\documentclass[danish,a4paper,10pt,twocolumn,amsmath,amssymb,floatfix,superscriptaddress,aps,pra,showpacs,longbibliography,groupedaddress]{revtex4-1}
\usepackage{graphicx}
\usepackage[english]{babel}
\usepackage{bm}
\usepackage{dcolumn}
\allowdisplaybreaks[4]

\newcommand{\fref}[1]{Fig.~\ref{#1}}

\newcommand{\be}{\begin{equation}}
\newcommand{\ee}{\end{equation}}
\newcommand{\ba}{\begin{eqnarray}}
\newcommand{\ea}{\end{eqnarray}}
\newcommand{\bs}{\begin{subequations}}
\newcommand{\es}{\end{subequations}}
\newcommand{\bw}{\begin{widetext}}
\newcommand{\ew}{\end{widetext}}

\usepackage{color}
\definecolor{red}{rgb}{1,0,0}
\definecolor{blue}{rgb}{0,0,1}
\usepackage{MnSymbol}

\begin{document}

\title{Light-Induced Structures in Attosecond Transient Absorption Spectroscopy of Molecules}

\author{Jens~E.~B\ae kh\o j}

\author{Lars~Bojer~Madsen}
\affiliation{Department of Physics and Astronomy, Aarhus University, 8000 Aarhus C, Denmark}

\date \today
\begin{abstract}
The nature of light-induced structures in attosecond transient absorption spectroscopy of molecular systems is investigated theoretically. It is shown how nuclear dynamics affect these structures. We find that a theoretical three-surface model captures the main characteristics in the calculated spectra. Based on this model, nuclear dynamics is divided into different categories, each category having unique signatures in the absorption spectra. 
Finally, we discuss the possibility for experimental observation of light-induced structures in molecules. 
\end{abstract}



\maketitle

\section{Introduction}

Since 2010 attosecond transient absorption spectroscopy (ATAS) has been used to study electronic dynamics in atoms on the attosecond time scale \cite{goulielmakis2010real, wang2010attosecond, santra2011theory, ott2014reconstruction}.
During propagation through a target region affected by an infrared (IR) or a near-infrared (NIR) pulse, the spectral profile of the attosecond pulse is  modified. After propagation the intensity spectrum is recorded to measure the accumulated modification. 
This spectrum depends on the relative phase between the attosecond pulse and the NIR/IR pulse and therefore on the time delay between the two pulses. 
Indirect temporal information about the response of the target to the NIR/IR field is therefore obtained from the delay-dependent intensity spectrum.
The spectral profile of the attosecond pulse is in the extreme ultraviolet (XUV) regime.
The intensity of the XUV pulse is low, and only processes involving a single XUV photon are non-negligible. 
The intensity of the NIR/IR pulse is in a regime where few-photon processes occur, but the intensity is too low to give rise to a signal in the XUV regime of interest.

Of particular interest in ATAS is the time interval where the XUV and the NIR/IR pulses overlap. In this region a number of spectral features arise in the spectrum as a consequence of non-linear coupling between the system and the two fields \cite{chen2012light, pfeiffer2012transmission, wu2013time, chen2013quantum}.
Here we focus on light-induced structures (LISs).
LISs are signatures of non-linear processes transferring population from the ground state to a state of the same parity (a dark state) by absorbing an XUV photon and absorbing or emitting a NIR/IR photon. 
In atomic systems LISs are therefore seen in ATAS spectra as pairs with each individual absorption feature located one NIR/IR photon energy above (emission of NIR photon) or below (absorption of NIR photon) the dark state energies.
The spectral width of an atomic LIS is inversely proportional to the duration of the NIR/IR pulse and since this pulse is often almost as long as the natural damping of the system, LISs appear as relatively narrow features in the ATAS spectrum \cite{chen2012light,beck2015probing}.
In Ref.~\citep{Egen} it was shown that molecular ATAS spectra are to a large extend determined by molecular vibrations,
and the results indicated that LISs at well-defined energies are not a general molecular property.
It is therefore of interest to make a careful investigation of how the LISs are affected by nuclear motion. 
It turns out that molecular LISs are extremely sensitive to the the type of nuclear dynamics supported by the electronic dark state Born-Oppenheimer (BO) curve.     
LISs therefore lend themselves to a study of nuclear dynamics in a new setting and they form a possibility to study the combined nuclear and electronic dynamics in the otherwise hidden dark states. The main purpose of this work is to give a more complete description of LISs in molecules, and show how LISs can be used to classify molecular processes. 

The paper is organized as follows.
Section~\ref{Theory} introduces the theoretical model used to obtain the ATAS spectra, the systems, and parameters used in the calculations.
Further, we introduce a three-curve model to isolate the main physical mechanics of LISs in molecules.
In Sec.~\ref{RES} we first show LISs in the ATAS spectrum of ${\text{H}_2}^+$ for fixed and moving nuclei. In ${\text{H}_2}^+$ all excited BO curves are dissociative. 
To investigate the effect of bound nuclear motion on ATAS spectra in general, and on LISs in particular, we consider classes of generic potential energy curves.
The resulting ATAS spectra are simple enough that the characteristics of LISs can be identified clearly.
Section~\ref{CON} concludes the paper and relates the results to a possible experimental observation of LISs in molecules.
Atomic units are used throughout.

\section{Models} \label{Theory}
To calculate the delay-dependent ATAS spectrum we use the method of Ref.~\cite{baggesen2012theory} (see also Ref.~\citep{Egen} where this method was related to other approaches). Assuming that  effects accumulated in propagation though the target region can be captured in a single-system response model \cite{gaarde2011transient}, the function 
\begin{align}
\tilde{S}(\omega,\tau)=2\text{Re}\left[ E_\text{in}^*(\omega,\tau) E_\text{gen}(\omega,\tau) \right],
\label{response}
\end{align}
gives the modification of the intensity spectrum at angular frequency $\omega$ and time delay $\tau$.
In Eq.~\eqref{response}, $E_\text{in}(\omega,\tau)$ is the Fourier transform of the unmodified, incoming field $E_\text{in}\left( t-\frac{x}{c},\tau \right)$
\begin{align}
E_\text{in}(\omega,\tau)&=\frac{1}{2\pi}\int_{-\infty}^{\infty} E_\text{in} \left( t-\frac{x}{c} , \tau \right) \exp(i\omega t) dt
\label{F_E_in}
\end{align}
and $E_\text{gen}(\omega,\tau)$ is the Fourier transform of the field $E_\text{gen}(x,t,\tau)$ generated by the system as a response to the incoming field \cite{baggesen2012theory}
\begin{subequations}
\begin{align}
E_\text{gen}(\omega,\tau)&=\frac{1}{2\pi}\int_{-\infty}^{\infty} E_\text{gen} \left( x,t,\tau \right) \exp(i\omega t) dt \label{pre_F_E_gen} \\
&=\frac{\omega i n}{c} \int_{-\infty}^{\infty} \langle d(t,\tau) \rangle \exp(i\omega t) dt,
\label{F_E_gen}
\end{align}
\end{subequations}
where $n$ is the target density, $c$ is the speed of light, and $\langle d(t,\tau) \rangle$ is the expectation value of the single-system dipole moment, found from quantum mechanical calculations.

We consider fields linearly polarized along the $z$-direction, and propagating in the positive $x$-direction. The $z$-component of the electric field $E(t)=-\partial_t A(t)$ used to model the XUV and NIR/IR pulses is defined through the vector potential
\begin{align}
A(t)=A_0 \exp \left[ -\frac{(t-t_c)^2}{T^2/4} \right] \cos \left[ \omega (t-t_c) \right], \nonumber \\ \quad T=N_c T_c=N_c\frac{2\pi}{\omega} ,
\label{vec_pot}
\end{align}  
where $A_0=E_0/\omega$, and $E_0$ relates to the intensity as $I=\vert E_0 \vert^2$, with one atomic unit of intensity equal to $3.51 \times 10^{16}$ W/cm$^2$. In Eq.~\eqref{vec_pot} $t_c$ is the center of the pulse.
The full width at half maximum (FWHM) duration $T_\text{FWHM}$ is related to $T$ used in Eq.~\eqref{vec_pot} by $T_\text{FWHM}=\sqrt{\ln (2)} T$.
To model ${\text{H}_2}^+$ we consider a simplified one-dimensional (1D), soft Coulomb potential model where the molecular axis is aligned with the linearly polarized laser pulse. Within this model, electronic and nuclear degrees of freedom are treated exactly. For more information see Refs.~\cite{Yue13, yue2014dissociative}. To mimic experimental results we damp the dipole moment using a window function of a duration of 42 fs \citep{Egen}.

In Ref.~\citep{Egen} it was shown that the dynamics in ${\text{H}_2}^+$ of importance for ATAS are well-described by an expansion in a few electronic states. In this work we therefore expand the wave functions in the lowest $N$ electronic BO states, $\phi_j(z;R)$: 
\begin{align}
\Psi (z,R,t) = \sum_{j=1}^N  G_j(R,t) \phi_j(z;R),
\label{state_N}
\end{align}
where $G_j(R,t)$ is the nuclear wave packet corresponding to the electronic state $\phi_j(z;R)$, $R$ is the internuclear separation, and $z$ denotes the electronic coordinate. 
For the analysis of nuclear dynamics it is sometimes useful to further expand the nuclear wave packets $G_i (R,t)$ in vibrational eigenstates $\chi_{i,k}(R)$ (discrete states) and $\chi_{E_i}(R)$ (nuclear continuum states) corresponding to the $i$'th electronic state
\begin{align}
G_i(R,t)=& \sum_k c_{i,k} (t) \chi_{i,k} (R) e^{-iE_{i,k}t} \nonumber \\ +&\int dE_i \; c_{E_i}(t) \chi_{E_i}(R) e^{-iE_i t}  \nonumber \\
\equiv &\sumint_k dE_{i,k} \; c_{i,k} (t) \chi_{i,k} (R) e^{-i E_{i,k} t},
\label{exp_j}
\end{align}
where $E_{i,k}$ and $E_{i}$ are the discrete and the continuous eigenenergies of the vibrational states of the $i$'th electronic state. Note that all vibrational states corresponding to a specific electronic state can be nuclear-continuum states (see Ref.~\citep{Egen} for details).

LISs found in ATAS spectra are a consequence of the NIR/IR coupling between an excited state of parity opposite from the ground state (bright state) and an excited state with the same parity as the ground state; the dark state.
The latter state is refereed to as a dark state since the transition from the ground state to this state is dipole forbidden and there are therefore no spectral features in the dipole response corresponding to the energy difference between dark states and the ground state.
Consequently these states are not observed in one-photon spectroscopy.
In Ref.~\cite{chen2012light} it was shown that the main characteristics of LISs in atoms are often captured in calculations using a three-state model. Even though LISs and the effect of nuclear motion on LISs was not the main focus of Ref.~\citep{Egen}, results of that paper indicated that LISs in molecules are also often well-described in a model only including three electronic states (three-curve model).
In the three-curve model the ground state $\phi_g(z,R)$, the bright state $\phi_e(z,R)$, and the dark state $\phi_d(z,R)$ [see Eq.~\eqref{state_N}] are included. Within this model a general state can therefore be written as
\begin{align}
\Psi(z,R,t)=& G_g(R,t) \phi_g (z;R) + G_d(R,t) \phi_d (z;R) \nonumber \\ +& G_e(R,t) \phi_e(z;R) \nonumber \\
=&\chi_{g,0}(R)\phi_g (z;R) \nonumber \\ +& \sumint_l dE_{d,l} c_{d,l}(t) \chi_{d,l} (R) e^{-iE_{d,l}(t-\tau)} \phi_d (z;R) \nonumber \\ +& \sumint_k dE_{e,k} c_{e,k}(t) \chi_{e,k}(R) e^{-iE_{e,k}(t-\tau)} \phi_e (z;R). \label{Nuc_3_state}
\end{align}
In obtaining the second part of Eq.~\eqref{Nuc_3_state} it is used that the ground state to a good approximation remains unchanged during propagation.
To analyze LISs in molecular ATAS spectra we also use the perturbative limit of the three-curve model: Assuming that the electric field is monochromatic [$E(t)=E_0 \sin (\omega t)$] we obtain that the first-order change in the expansion coefficient $c_{e,k}(t)$, for the $k$'th vibrational state of the bright electronic state, is given by 
\begin{align}
c_{e,k}^{(1)}(t) =& \frac{E_0  d_{d,e}^\text{el} (R_0)}{2i} \sumint_l dE_{d,l} \nonumber \\ &\bigg[ \frac{e^{i(E_{e,k}-E_{d,l}+\omega)\times(t-\tau)}-1}{E_{e,k}-E_{d,l}+\omega}e^{i\omega\tau} -  \nonumber \\ & \frac{e^{i(E_{e,k}-E_{d,l}-\omega)\times(t-\tau)}-1}{E_{e,k}-E_{d,l}-\omega}e^{-i\omega\tau} \bigg] \langle \chi_{e,k} \vert \chi_{d,l} \rangle c_{d,l}(\tau).
\label{dipole_three_nuc_coef}
\end{align}
The expectation value of the dipole moment in the three-curve model of Eq.~\eqref{Nuc_3_state} is well approximated by
\begin{align}
\langle d(t,\tau) \rangle \simeq & 2 \text{Re} \bigg[ d_{g,e}^\text{el}(R_0) \sumint_k dE_{e,k} \nonumber \\ &\left\langle \chi_{g,0}  \vert \chi_{e,k} \right\rangle \; c_{e,k} (t) e^{-iE_{e,k}(t-\tau)} \bigg]. \label{dipole_three_nuc}
\end{align}
To obtain Eq.~\eqref{dipole_three_nuc} it is assumed that the electronic dipole moment function
\begin{align}
d^{\, \text{el}}_{i,j} (R)=\int dz \; \phi_i^* (z;R) d \phi_j (z;R)
\label{el_dip}
\end{align}
is a slowly varying function of $R$ and therefore can be evaluated at $R_0=\int R \, \vert G_g(R,t=-\infty) \vert^2 dR$. This is usually a good approximation for ATAS calculations (see Ref.~\cite{Egen}).
Inserting Eq.~\eqref{dipole_three_nuc_coef} into Eq.~\eqref{dipole_three_nuc} shows that the energies $E_{e,k}$ disappear in the final expression for the time-dependent dipole moment, leaving terms oscillating with frequencies $E_{d,l} \pm \omega_c$. 
The distribution and population of vibrational states corresponding to the bright electronic state $\phi_e (z,R)$ does therefore not affect the dephasing of the dipole moment responsible for LISs.
The distribution and population of vibrational states corresponding to the dark electronic state $\phi_d (z,R)$ are, on the other hand, crucial for the nature of LISs. 
The more dark vibrational states that are populated, the faster the resulting dipole moment will dephase.
Even in the situation where a finite number of discrete vibrational states are populated, and one would expect a partial revival of the time-dependent dipole moment, the revival is often absent due to the finite duration of the NIR pulse. 
We therefore expect LISs with well-defined energies to be suppressed in the molecular ATAS spectrum when more than a few vibrational dark states are populated. The results in the next section verify this explanation.

\section{Results}
\label{RES}
In this section we present ATAS spectra calculated using Eq.~\eqref{response}. To find the time-dependent dipole moment we propagate the nuclear wave packets of Eq.~\eqref{state_N} in accordance with the time-dependent Schr\"{o}dinger equation (see Ref.~\cite{Egen} for details). 
First we calculate the ATAS spectra for our ${\text{H}_2}^+$ model with fixed and moving nuclei, then in Sec.~\ref{anden} we investigate how LISs depend on  the relative arrangement and shapes of the electronic BO curves.

\subsection{${\text{H}_2}^+$ model}
\begin{figure}
\includegraphics[width=0.42\textwidth]{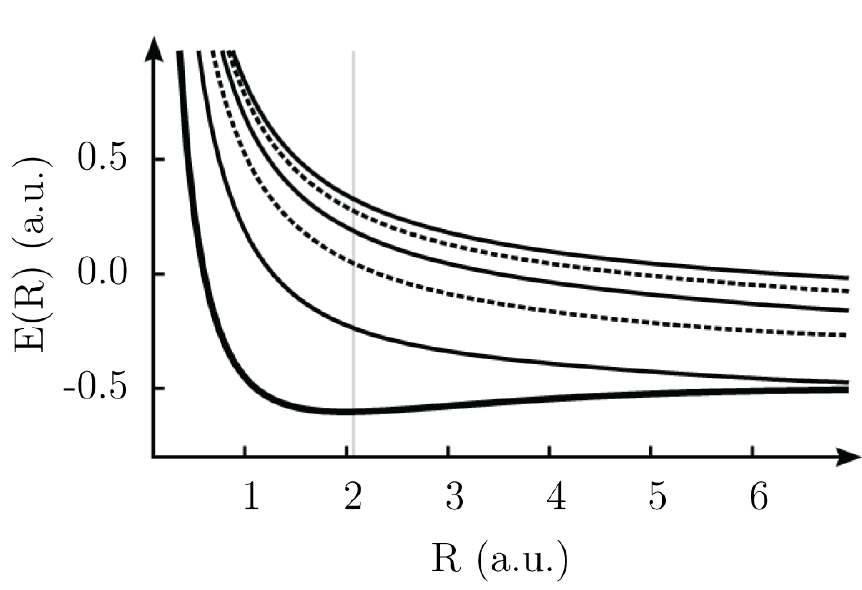}
\caption{\label{Fig1} BO curves of our ${\text{H}_2}^+$ model. The thick, full line shows the ground state, while the narrower full lines are bright excited states and the dashed lines are excited dark states (see text). The vertical gray line indicates the value of $R_0=\int R \, \vert G_g(R,t=-\infty) \vert^2 dR=2.07$.}
\end{figure}

\begin{figure}
\includegraphics[width=0.48\textwidth]{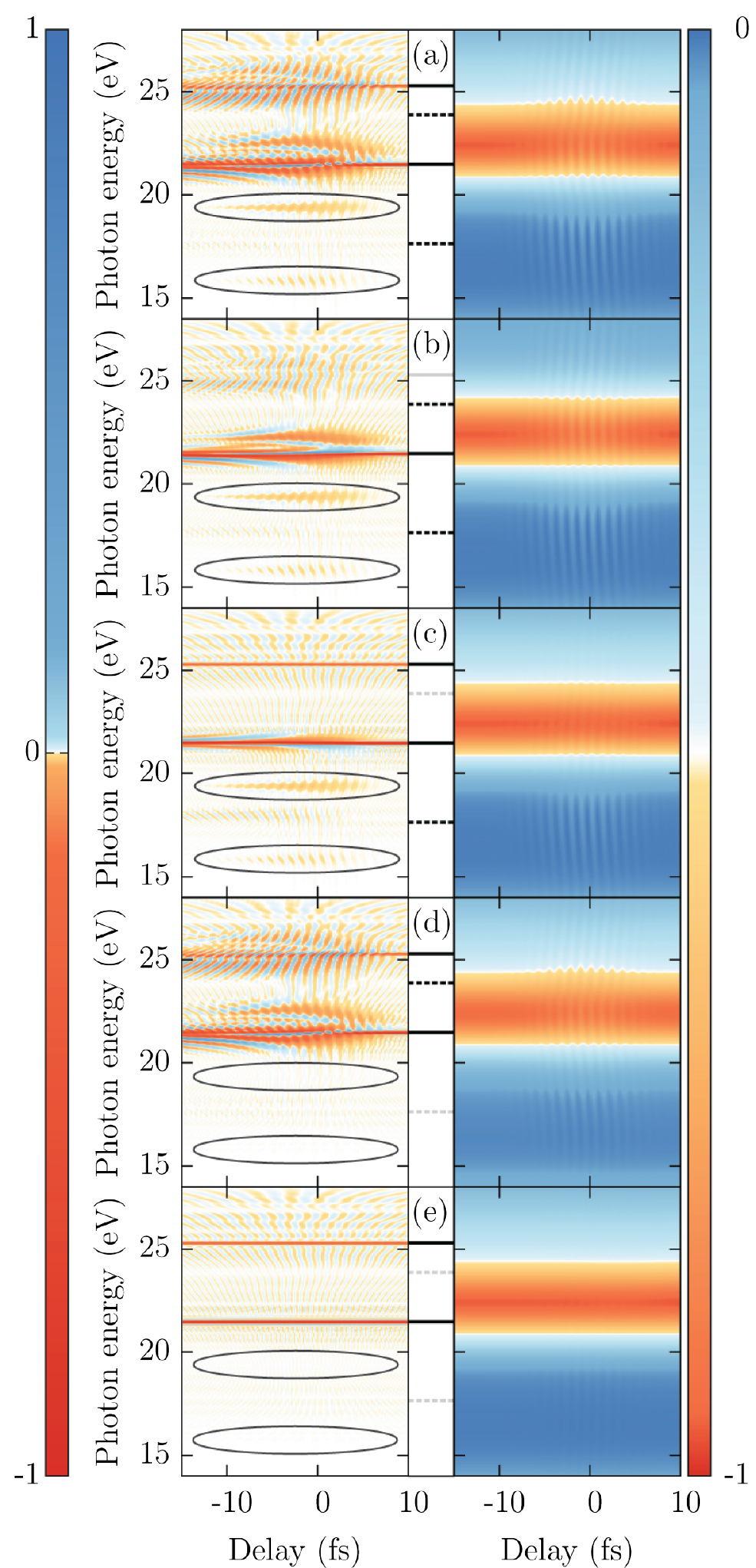}
\caption{\label{Fig2} (Color online) ATAS spectra $\tilde{S}(\omega,\tau)$ [Eq.~\eqref{response}] of ${\text{H}_2}^+$ with fixed (left column) and moving (right column) nuclei. (a) includes the six lowest electronic states in the calculation. The energy evaluated at $R_0$ of the four states in the energy range of interest are shown in the middle column. In the middle column full lines represent bright states and dashed lines represent dark states (see text). (b)-(e) excludes the electronic states indicated by light-gray coloring in the middle column. Note that there are separate color bars for the left and right columns. The color bars are designed to highlight the LISs which are also encircled by black cigars in the fixed nuclei case. 
 Pulse parameters: $T_\text{XUV}=330$ as, $T_\text{NIR}=14.0$ fs, $\lambda_\text{XUV}=50$ nm, $\lambda_\text{NIR}=700$ nm, $I_\text{XUV}=5\times 10^7$ W/cm$^2$, $I_\text{NIR}=2\times 10^{12}$ W/cm$^2$.}
\end{figure}

In \fref{Fig1} we show the lowest six BO potential curves of our 1D model of ${\text{H}_2}^+$. In the model we use a softened Coulomb potential designed to reproduce the exact three-dimensional $1s\sigma_g$ BO curve \cite{Madsen12,Yue13}. The figure shows that a nuclear wave packet located on any excited BO curve dissociates.
In \fref{Fig2} the ATAS spectra are shown for the atomic-like situation with the nuclei fixed at $R=R_0$ (left panel) and for the molecular case including  nuclear motion (right panel). 
We immediately see from \fref{Fig2} that including nuclear motion in the description completely changes the ATAS spectra from atomic-like spectra characterized by narrow absorption lines to spectra dominated by very broad features.
In the following we will see that such broadening is characteristic, not only for spectral features due to dipole allowed transitions, but also for LISs.
We investigate how selected electronic states affect the spectrum by eliminating them from the basis set of the calculation. In this way the physical processes responsible for different spectral features are identified.

The ATAS spectra of \fref{Fig2}(a) are calculated using the expansion~\eqref{state_N} including the ground state and the five lowest excited electronic states. 
The energies of the four electronic states in the energy range of relevance are shown in the middle column of \fref{Fig2}(a), where energies corresponding to bright and dark states are indicated by full or dashed lines, respectively.
In Figs.~\ref{Fig2}(b)-(e) some of these lines are shown in light-gray coloring, indicating that the electronic state of that energy has been omitted in the calculations of the spectra. 
All energies are given with respect to the ground state energy.
The fixed nuclei spectrum [\fref{Fig2}(a), left] shows many similarities with atomic spectra. 
For large positive delays of the $T=330$ as XUV pulse [see Eq.~\eqref{vec_pot}] with respect to the NIR pulse the spectrum is characterized by narrow absorption features at the field-free energies of bright excited states at 21.48 eV and 25.29 eV (full lines in the middle column).
The LISs in the fixed nuclei spectrum, located one NIR-photon energy above/below the dark state of energy $17.65$ eV, are encircled by black cigar shapes and are characterized by droplet-like shapes, a periodic modulation with a period of half the period of the NIR cycle, and a width inversely proportional to the duration of the NIR pulse. 
These are all characteristics known from LISs in atomic systems \cite{beck2015probing,chen2012light}. We use a $T_\text{NIR}$=14 fs NIR pulse [see Eq.~\eqref{vec_pot}] and as a consequence our LISs have widths in the sub-eV domain. 
In the moving-nuclei spectrum the situation is very different [\fref{Fig2}(a), right]. All narrow absorption features are gone, and the two absorption lines corresponding to transitions to bright states (at $21.48$ eV and $25.29$ eV in the fixed-nuclei spectrum) have merged into a single very broad feature. 
More important for the current discussion is that the LISs at the relatively well-defined energies of $\sim 16$ and $\sim 19$ eV appear to be absent in the moving-nuclei spectrum.
The only visible NIR-field modulation of the moving-nuclei spectrum in the right panel of \fref{Fig2}(a) is a periodic modulation with a period of half a NIR-field cycle from $-5$ fs to $5$ fs, extending over the entire energy range of the figure. 
Since this modulation seems to be centered around the broad absorption line at $\sim 21-24$ eV it is natural to assume that this modulation is a consequence of 'which-way interference' \cite{chen2013quantum} between the two bright states [at $21.48$ eV and $25.29$ eV, indicated by full lines in the middle column in \fref{Fig2}(a)]. 
In \fref{Fig2}(b) we calculate the ATAS spectrum again, but this time without the highest excited bright state (now a gray line at $25.29$ eV in the middle column). 
Since only one bright state is left in the frequency domain of interest, we expect all modulations caused by which-way interference to be eliminated. 
From the moving-nuclei spectrum of \fref{Fig2}(b), however, we see that the spectrum is remarkably unaffected by the removal of the bright state. 
For the fixed-nuclei spectrum of \fref{Fig2}(b) we see that the LISs at $\sim 16$ eV and $\sim 19$ eV are also unaffected. 
Since the interference between bright states apparently can not explain the modulation of the moving-nuclei ATAS spectrum we are led to expect that this modulation is caused by LISs. 
To validate this expectation we calculate the ATAS spectrum without the highest excited dark state [at 23.87 eV, gray dashed line of \fref{Fig2}(c)] and without the second highest excited dark state [at 17.65 eV, gray dashed line of \fref{Fig2}(d)]. 
Without the dark states there will not be any LISs, and from the fixed-nuclei spectrum of \fref{Fig2}(d) we see that the LISs at $\sim 16$ eV and $\sim 19$ eV are indeed missing when the $17.65$ eV dark state is removed from the basis set of the calculation. 
In the moving-nuclei spectra we see that removal of the highest excited dark state reduces the modulation in the high energy part of the spectrum [\fref{Fig2}(c)], and removal of the lower excited dark state reduces the modulation in the low energy part of the spectrum [\fref{Fig2}(d)]. 
Finally in \fref{Fig2}(e) we remove both dark states from the calculation. In the corresponding fixed and moving-nuclei spectra the effect of the NIR pulse is almost gone. 
It is not surprising that \fref{Fig2}(e) is almost independent of the NIR pulse since all dark states have been removed. When \fref{Fig2}(e) is hold together with Figs.~\ref{Fig2}(b)-(d) it is, however, clear that the NIR-modulations in the moving nuclei spectrum of \fref{Fig2}(a) are primarily resulting from LISs.

Returning to \fref{Fig2}(c) we observe that the removal of the highest excited dark state causes a drastic reduction in the $T_\text{NIR}/2$ oscillations in the absorption lines at $21.48$ eV and $25.29$ eV. These oscillations are a result of which-way interference, and we therefore conclude that the dark state is essential in the coupling of the two bright states which is also anticipated from, e.g., second-order perturbation theory.
Additionally it is seen from \fref{Fig2}(c) that the absorption feature at $\sim 23$ eV stretching from $\sim -10$ to $\sim 5$ fs [in Figs.~\ref{Fig2}(a), (b) and (d)] disappears from the spectrum. This absorption feature is therefore also a LIS.

We have argued that the periodic modulation of the moving-nuclei spectrum of ${\text{H}_2}^+$ [\fref{Fig2}(a), right], stretching over a broad energy range, are in fact LISs. To explain why these LISs do not look anything like what we see in atomic systems and as in ${\text{H}_2}^+$ with fixed $R$ we return to Eqs.~\eqref{dipole_three_nuc_coef} and \eqref{dipole_three_nuc}.
From these equations we were able to conclude that LISs would be broadened if more dark vibrational states were populated since this leads to a dephasing of the part of the time-dependent dipole moment responsible for LISs.
Based on the Frank-Condon principle we expect that the nuclear wave function, excited to a dark state BO curve, will initially be localized around $R=R_0$, but since all dark states in ${\text{H}_2}^+$ are dissociative, all the corresponding vibrational states are nuclear-continuum states. 
An expansion of the nuclear wave packet in vibrational dark states will therefore in principle require an infinite number of dark vibrational states. 
As a result we expect that the corresponding dipole signal will be damped on a very short time scale leading to extremely broad spectral features. 
How broad the LISs are depends on the distribution of vibrational dark states populated by the combined XUV and NIR pulses and therefore on the slope of the corresponding BO curve.

\subsection{Illustrative examples using the three-curve model}
\label{anden} 
\begin{figure}
\includegraphics[width=0.42\textwidth]{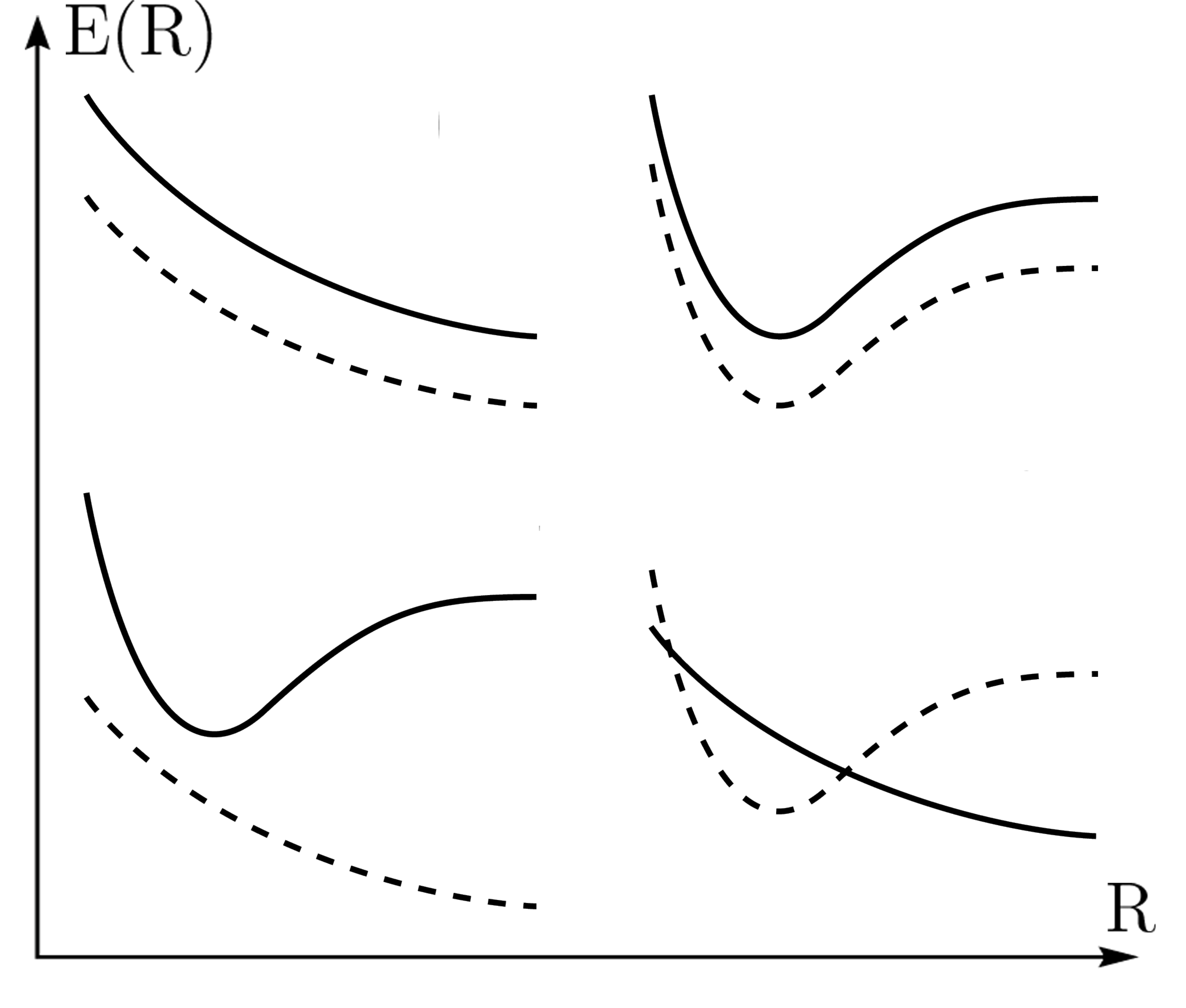}
\caption{\label{Fig3} (Color online) Illustration of possible arrangements of excited BO curves in molecules. Full lines represent bright states with respect to coupling from the ground state curve, while dashed lines represent dark states.}
\end{figure}

In many situations LISs can be accurately described using a three-curve model containing the electronic ground state, one electronic dark state and a single electronic bright state [see Eq.~\eqref{Nuc_3_state}]. 
We take the ground state BO curve to be binding. For the excited curves, we consider different shapes. 
Our goal is to investigate how the shapes of the two excited BO curves in the three-curve model affect the LISs. We divide the excited BO curves into two categories: binding and dissociative. 
This division gives us four different characteristic arrangements of BO curves as shown schematically in \fref{Fig3}. 
The upper left set of BO curves in \fref{Fig3} represents the case where both excited BO curves are dissociative. 
This case is similar to ${\text{H}_2}^+$ discussed in the previous section. 
We have confirmed that the three-curve model is capable of reproducing the LISs in the ATAS spectra of \fref{Fig2}, corresponding to the included dark state, for both fixed and moving-nuclei calculations.
To obtain the other arrangements of BO curves shown in \fref{Fig3} we replace one or two of the excited BO curves $C_d(R)$ (dashed, dark state curve) and $C_e(R)$ (full, bright state curve) of ${\text{H}_2}^+$ with the curves
\begin{align}
\tilde{C}_{d/e}(R)=C_g(R-R_{d/e})-C_g(R_0)+C_{d/e}(R_0),
\label{arti_curve}
\end{align}
where $C_g(R)$ is the ground state curve. 
For $R_{d/e}=0$ the curve $\tilde{C}_{d/e}(R)$ is parallel to the ground state and the energy difference to the ground state curve is as for the original curve $C_{d/e}(R)$ evaluated at $R_0$. 
In the numerical implementation of the model we have used the dipole matrix elements from the ${\text{H}_2}^+$ case. 
We have found that the main characteristics of LISs in the spectra of the three-curve model are not largely affected by the specific choice of dipole matrix element values, when reasonable values are chosen.

\begin{figure}
\includegraphics[width=0.48\textwidth]{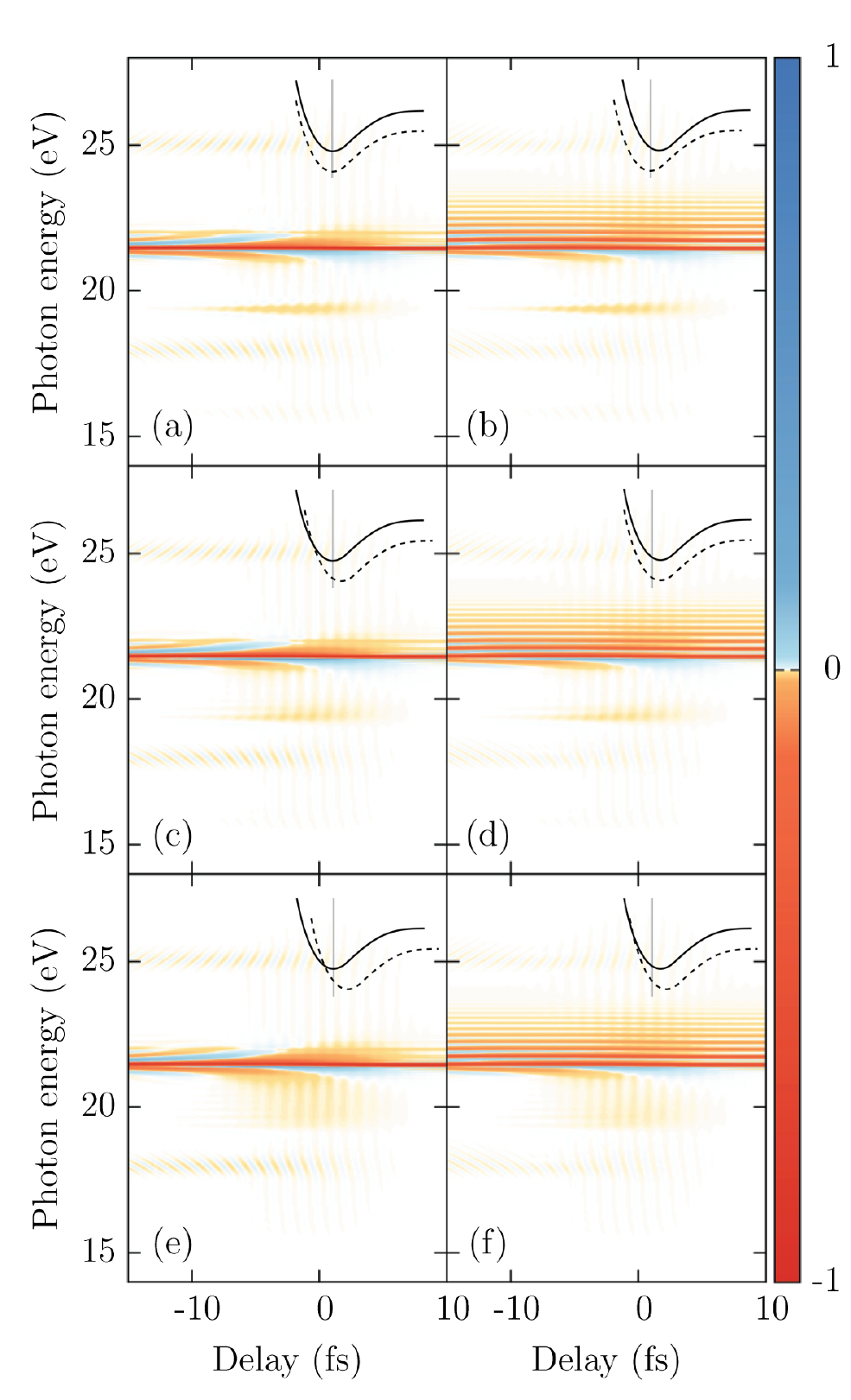}
\caption{\label{Fig4} (Color online) ATAS spectra of a model with two binding excited BO curves [see upper right part of \fref{Fig3}]. The spectra are calculated for the different values of $R_d$ and $R_e$ [see Eq.~\eqref{arti_curve}] $R_d=R_e=0.0$ (a), $R_d=0.0$ and $R_e=0.3$ (b), $R_d=0.3$ and $R_e=0.0$ (c), $R_d=0.3$ and $R_e=0.3$ (d), $R_d=0.5$ and $R_e=0.0$ (e), and $R_d=0.5$ and $R_e=0.3$ (f). Special attention should be given to the spectral feature in (a) and (b) at $\sim 19$ eV. The insets show illustrations of the two excited BO curves included in the calculations. In the insets, the vertical gray line indicate $R=R_0$. Pulse parameters are as in \fref{Fig2}.}
\end{figure}

The upper right arrangement of BO curves in \fref{Fig3} represent the case where both excited curves are binding. 
In \fref{Fig4} we show the corresponding ATAS spectra for different values of $R_d$ and $R_e$, i.e., for different relative positions of the bright and dark state BO energy curves.
In \fref{Fig4}(a) ($R_d=R_e=0$) we see a narrow absorption line at $\sim 22$ eV for the dipole allowed transition and a narrow single-line LIS at $\sim 19$ eV. 
This atomic-like behavior is expected since all three BO curves are parallel. 
The vibrational ground state is therefore the same in all three curves, and from the Franck-Condon principle nuclear dynamics does not play any important role, since excited vibrational states are never populated. 
In \fref{Fig4}(b) the minimum of the bright state BO curve is at larger $R$ [Eq.~\eqref{arti_curve} with $R_e=0.3$]. 
When the lowest vibrational state of the electronic ground state is expanded in the (discrete) vibrational states of the bright state curve several of the expansion coefficients will be non-zero. 
In the spectrum we therefore see several absorption lines corresponding to the energy of the individual vibrational states. 
The LIS at $\sim 19$ eV in \fref{Fig4}(b), however, are remarkably unaffected by the translation of the bright state curve. This is in agreement with the conclusion in Sec.~\ref{Theory}, that the LISs are unaffected by the population and distribution of vibrational bright states. 
In Figs.~\ref{Fig4}(c) and \ref{Fig4}(d) the minimum of the dark state BO curve is moved to larger $R$ in the calculations ($R_d=0.3$). 
As a result several vibrational dark states are populated and we observe a broadening of LISs in the spectrum. 
A careful examination of Figs. \ref{Fig4}(c) and \ref{Fig4}(d) shows that the LIS at $\sim 19$ eV has a horizontal few-line structure (not clearly visible at the scale of the figure) corresponding to the energies of the different populated vibrational dark states.
We therefore refer to this type of spectral features as few-line molecular LISs.
In Figs.~\ref{Fig4}(e) and \ref{Fig4}(f) the spectra for calculations with $R_d=0.5$ are shown, and we see even broader absorption lines without horizontal line structure (defuse LISs). 
In the latter case the slope of the dark BO curve evaluated at $R_0$ is steeper than for the $R_d=0.3$ case [Figs.~\ref{Fig4}(c) and \ref{Fig4}(d)] and it is therefore expected that more vibrational states become populated by the combined action of the XUV and NIR laser pulses.
The absorption features in Figs.~\ref{Fig4}(a)-(f) corresponding to dipole allowed transitions seems to be unaffected by the dark state. 
However, we know from the discussion of \fref{Fig2} that dark states play an important role in the transitions between bright states. 
In general, the shapes of dark state BO curves therefore affect the absorption lines corresponding to dipole allowed transitions.

\begin{figure}
\includegraphics[width=0.48\textwidth]{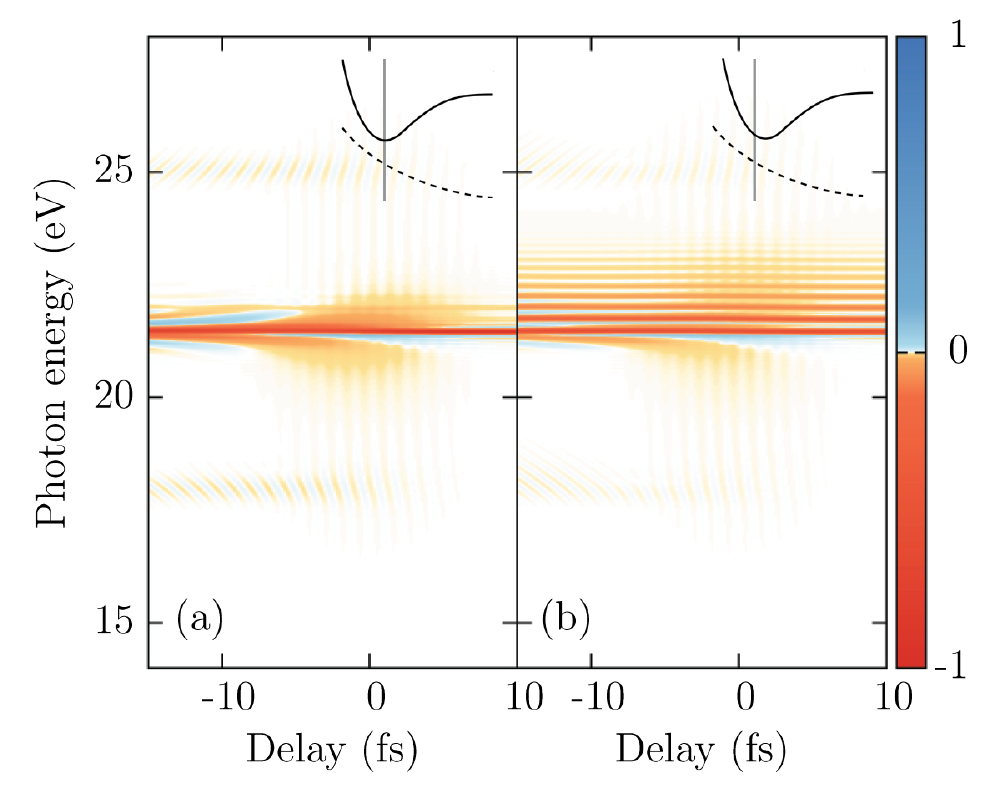}
\caption{\label{Fig5} (Color online) ATAS spectra of a model with a single  binding excited bright BO curve and a single dissociative dark BO curve [see lower left part of \fref{Fig3}]. In the calculations $R_e$ of Eq.~\eqref{arti_curve} is chosen as $R_e=0$ (a) or $R_e=0.3$ (b). The insets show illustrations of the two excited BO curves included in the calculations. In the insets, the vertical gray lines indicate $R=R_0$. Pulse parameters are as in \fref{Fig2}.}
\end{figure}

We now investigate the case of a bound bright BO curve and a dissociative dark BO curve schematically drawn in the lower left part of \fref{Fig3}. 
The corresponding ATAS spectra in \fref{Fig5} show that absorption lines resulting from dipole allowed transitions are again very similar to what we saw in \fref{Fig4}.
The LISs, however, are very defuse, stretching over a broad energy region. This is similar to what we saw in the moving-nuclei spectra of ${\text{H}_2}^+$ shown in the right panel of \fref{Fig2}(a). 
The fact that the LISs of \fref{Fig5} are similar to LISs in the moving nuclei spectrum of ${\text{H}_2}^+$ is another example that often the characteristics of LISs do not depend on the shape of the bright state curve.

\begin{figure}
\includegraphics[width=0.48\textwidth]{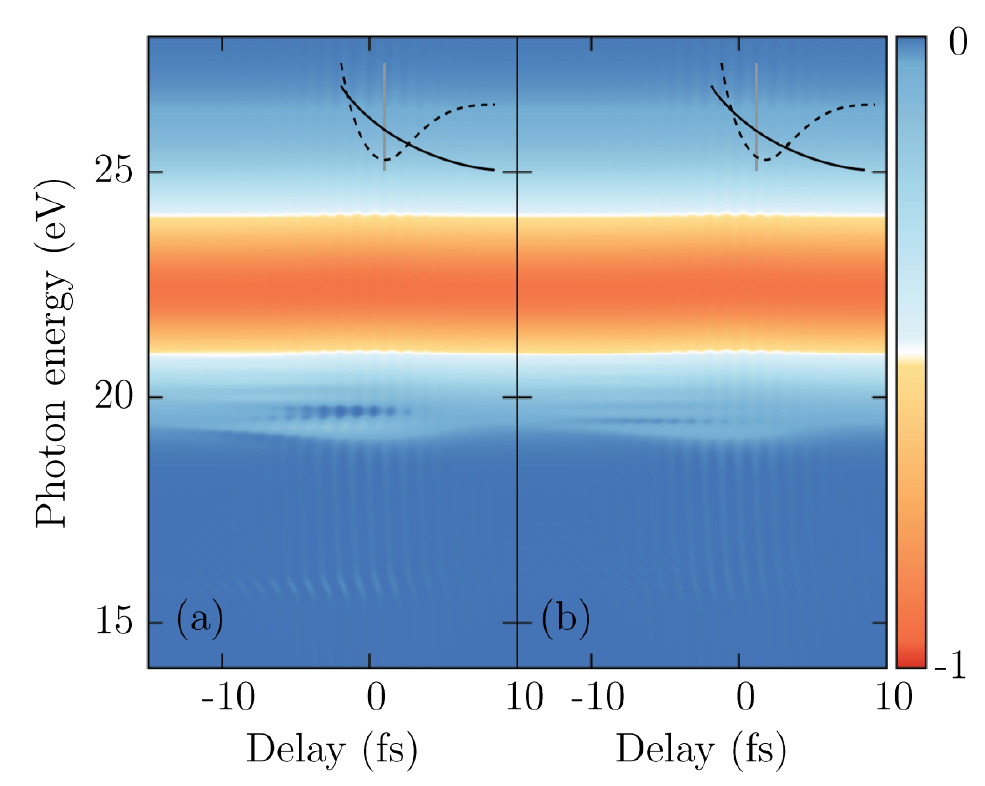}
\caption{\label{Fig6} (Color online) ATAS spectra of a model with a single dissociative excited bright BO curve and a single binding dark BO curve [see lower right part of \fref{Fig3}]. In the calculations $R_d$ of Eq.~\eqref{arti_curve} is chosen as $R_d=0$ (a) or $R_d=0.3$ (b). The insets show illustrations of the two excited BO curves included in the calculations. In the insets, the vertical gray lines indicate $R=R_0$. Pulse parameters are as in \fref{Fig2}.}
\end{figure}

In \fref{Fig6} we show the spectra of a molecule with a bound dark state curve and a dissociative bright state curve as sketched in the lower right corner of \fref{Fig3}. 
In \fref{Fig6}(a) we see that there are several absorption features above the main LIS at $\sim 19$ eV [also seen in, e.g., \fref{Fig4}(a)]. 
Since the ground state and dark state curves are parallel, this structure indicates that the LISs are in fact affected by the bright state curve in this case. 
For the calculations carried out to produce \fref{Fig6}(a) the slope of the dark state BO curve evaluated at $R_0$ is so steep that the nuclear wave function $G_e(R,t)$ slide down the curve on a timescale comparable to the duration of the XUV pulse. 
As a result several vibrational dark states are populated by the combined action of the XUV and NIR fields even though the ground state and dark state curves are parallel.
In spectra for models where a less steep dissociative bright state is used, the LISs are again unaffected by the shape of the bright state curve.
For the calculations producing \fref{Fig6}(b) the minimum of the dark state curve has been moved to larger $R$-values, but the spectra of Figs.~\ref{Fig6}(a) and \ref{Fig6}(b) are very similar. 
In contrast to earlier cases it is difficult to say much about the position and shape of the dark BO curve from LISs when the bright state curve is very steep compared to the dark state curve.
The effect described above can be minimized by using a shorter XUV pulse.

\section{Conclusion and discussion}
\label{CON}
We have investigated LISs in four different arrangements of excited state BO curves [see \fref{Fig3}]. 
We found that the characteristics of LISs are extremely dependent on the shape of the dark state BO curve which is not directly dipole coupled with the ground state. Conversely, the LISs are in many cases independent of the shapes of the bright, dipole allowed BO curves.
LISs can therefore be used as a measure of the dark state BO curve-shape. 
A narrow single-line molecular LIS is a signature of a dark BO curve parallel or almost parallel to the ground state [see \fref{Fig4}(a)], while non-parallel curves and curves with shapes different from the ground state curve give horizontal few-line or defuse LISs. 
We see a continuous change in the molecular LISs from narrow to defuse, when more and more vibrational dark states are needed in the expansion of the nuclear wave function of the ground state. 
The impact of nuclear motion can therefore be divided into characteristic categories including single-line molecular LISs when the expectation value of the internuclear separation on the dark state curve $\int R \, \vert G_d (R,t) \vert^2 dR$ is constant in time and defuse LISs when the dark state curve is dissociative.

In addition to the results shown in the present work, we have performed calculations on the 3D $\text{H}_2$ model of Ref.~\citep{Egen}. It was, however, difficult to unambiguously identify LISs in the spectra. 
There are several reasons for this.
First, the only dark state curve (EF) in the energy range of interest is very different from the ground state curve (X) (For information about the BO curves in H$_2$ see Ref.~\citep{fantz2006franck}). 
The ground state curve is a single-minimum curve with a minimum at $R=1.37$ while the dark state EF curve has two minima at $R=1.91$ and $R=4.41$, respectively. 
As a consequence the overlap between vibrational ground states of the two BO curves are much smaller than one, and we expect the corresponding LISs to be very diffuse. 
In contrast to the ATAS spectra of this paper, the spectra of H$_2$ are very rich in structures (see Fig.~(5) in Ref.~\cite{Egen}). 
Two bright state BO curves (B and C) are located in the energy range where we expect to see LISs and as a consequence it is nearly impossible to distinguish LISs from, e.g., which-way interference phenomena.
To see clear LISs the dipole matrix elements between the ground state and a bright state and between this bright state and a dark state should both be large [see Eqs.~\eqref{dipole_three_nuc_coef} and \eqref{dipole_three_nuc}] compared to dipole matrix elements describing processes responsible for other spectral features in the energy region of the LISs. 
In H$_2$ unambiguous identification of LISs is hindered by which-way interference between vibrational states of the B and C curves since the dipole matrix elements are almost ideal for this process.
Tuning the coupling strengths of the system as outlined above, to circumvent the difficulties of H$_2$, we observe LISs behaving very similarly to what we saw in Sec.~\ref{RES}.
We therefore expect that also in experiments LISs may in many cases be observed and used to divide nuclear dynamics on dark BO curves into categories: stationary when single-line LISs are observed, small amplitude oscillations when horizontal few-line LISs are observed and large amplitude oscillations or dissociative behavior when diffuse LISs are observed.

\section*{Acknowledgments}
This work was supported by the ERC-StG (project no. 277767-TDMET), and the VKR center of excellence QUSCOPE.


%

\end{document}